\documentstyle[aps,prl,epsf,twocolumn]{revtex}

     \def\@cite#1{{\footnotesize $^{#1}$}}
\begin{document}
\draft
\twocolumn[\hsize\textwidth\columnwidth\hsize\csname @twocolumnfalse\endcsname
\title{ Mean Field Theory of The Mott-Anderson Transition\\
{\normalsize\rm 
[Journal reference: Phys. Rev. Lett. {\bf}78, 3943 (1997)]}}
\author{ V. Dobrosavljevi\'{c}}
\address{Department of Physics and 
National High Magnetic Field Laboratory, \\Florida State University, 
Tallahassee, Florida 32306.}
\author{ G. Kotliar}
\address{Serin Physics Laboratory,
Rutgers University,
PO Box 849,
Piscataway NJ, 08855.
}
\maketitle

\begin{abstract}

We present a theory for disordered interacting electrons that 
can describe both the Mott and the Anderson transition in the 
respective limits of zero disorder and zero interaction. We use it   
to investigate  the $T=0$ Mott-Anderson transition at a fixed electron
density, as a the disorder strength is increased.
Surprisingly, we find  {\em  two}
critical values of disorder $ W_{nfl}$ and  $W_c$.
For $W > W_{nfl}$, the system enters a 
``Griffiths'' phase, displaying metallic non-Fermi liquid behavior. 
At even stronger disorder, $W=W_c >  W_{nfl}$ the system undergoes a 
metal insulator transition, characterized by the 
linear vanishing of both the {\it typical} density of states and 
the {\it typical} quasiparticle weight. 

\end{abstract}

\pacs{PACS Numbers: 75.20.Hr, 71.55.Jv}  
]
\narrowtext

The nature of the metal-insulator transition, is a fundamental
problem in condensed matter science.
There are two basic mechanisms that cause electron localization. 
Mott demonstrated  that electron-electron 
interactions,  can produce a metal insulator transition (MIT)
even in a clean system \cite{mott}. Anderson discovered  that 
disorder, i. e. strong spatial fluctuations in the potential 
due to impurities \cite{anderson}, 
can drive a metal insulator transition in a system of non interacting
electrons.

Following these early ideas, important advances were made following
the application of scaling approaches 
\cite{gang4,wegner,efetov,fink,ckl,bk,tednew,ted} to the problem.
In the interacting case, these
formulations turned out to be closely connected to
Fermi liquid ideas \cite{ckl}. 

These efforts notwithstanding,  many basic questions remain.
In particular, it  
proved very difficult to  incorporate the effects of strong electronic
correlations,  such as the formation of local magnetic moments,
in a comprehensive theory of the MIT. 
This is a serious shortcoming, since it is 
well established  experimentally that
the metallic state close to 
the MIT is characterized by a divergent magnetic
susceptibility and linear specific heat coefficient. These observations
form the basis of the two fluid phenomenology \cite{twoflu}. 

Very recently, a new approach \cite{dinf} to the strong correlation
problem has been developed and successfully applied to systems in the
vicinity of the Mott transition.  This dynamical mean-field theory is
in its spirit quite similar to the well known Bragg-Williams theory of
magnetism, and as such becomes exact in the limit of large
coordination.  The approach has furthermore been extended to
disordered systems \cite{dinfdis}, and used to investigate phenomena
such as disorder-induced local moment formation \cite{msb}.  However,
if formulated in its strict large-coordination limit, the theory
misses strong spatial fluctuations, and thus cannot incorporate
Anderson localization effects.

The goal of the present study is to present a theory that can
describe both the Mott and the Anderson route to localization, 
and therefore  address the interplay of these effects. We follow an 
approach very similar to the well known Thouless-Anderson-Palmer (TAP)
formulation of the mean field theory of spin glasses \cite{tap}. 
Specifically, we treat the correlation aspects of the problem in a 
dynamical mean-field theory fashion, but allow {\em spatial variations}
of the order parameter in order to allow for Anderson localization effects.
The theory is then exact in the noninteracting limit, and reduces to the
standard dynamical mean field theory in absence of disorder. 

For simplicity,
we consider a simple single-band Hubbard model with random site energies, 
as given by the Hamiltonian

\begin{eqnarray}
H&=&\sum_{ij}\sum_{\sigma } ( -t_{ij} + \varepsilon_i \delta_{ij})
c^{\dagger}_{i,\sigma}c_{j,\sigma}\nonumber +
U\sum_{i}c^{\dagger}_{i,\uparrow}c_{i,\uparrow}
c^{\dagger}_{i,\downarrow}c_{i,\downarrow}.
\end{eqnarray}

Within the dynamical mean-field theory, all local correlation functions 
can be evaluated using a single-site effective action of the form
\begin{eqnarray}
&&S_{eff} (i) =  \sum_{\sigma}\int_o^{\beta }d\tau\int_o^{\beta }d\tau '
c^{\dagger}_{i,\sigma} (\tau )( \delta (\tau -\tau ')\left(
\partial_{\tau} +\varepsilon_i -\mu \right) \nonumber \\
&& +  \Delta_{i,\sigma} (\tau,\tau ') )c_{i,\sigma} (\tau ')  
  + U\int_o^{\beta}d\tau n_{i,\uparrow}(\tau )n_{i,\downarrow}(\tau ).
\end{eqnarray}
Here, we have used functional integration over Grassmann  fields 
$c_{i,\sigma} (\tau )$ that represent electrons of spin
$\sigma$ on site $i$, and $n_{i,\sigma}(\tau )= 
c^{\dagger}_{i,\sigma} (\tau )c_{i,\sigma} (\tau )$.
The `` hybridization function''
 $\Delta_{i} (\tau,\tau ')$
 is obtained by formally integrating out all the degrees of freedom 
on  other sites in the lattice, and is given by
\begin{equation}
\label{eq3}
\Delta_{i} (\omega_n ) =\sum_{j=1}^z t_{ij}^2 G_{j}^{(i)} (\omega_n ).
\end{equation}
The sum over $j$ runs over the $z$ neighbors of the site $i$, and
$G_{j}^{(i)} (\omega_n )=<c^{\dagger}_{j}(\omega_n)c_{j} (\omega_n )>$ 
are the local Green's functions evaluated on 
site $j$, but with the site $i$ removed. 
For $z$ finite, and arbitrary 
lattices, $G_{j}^{(i)} (\omega_n )$ cannot be expressed through {\em local}
Green's functions only, but the situation is simpler on a Bethe 
lattice \cite{abouchacra}, where a simple recursion relation
can be written 
for this object, expressing it through similar objects on neighboring sites.
In particular, $G_{j}^{(i)} (\omega_n )$ can be computed from a local
action of the form  identical as in Eq. (2), except that in the 
expression for $\Delta_{j} (\tau,\tau ')$, the sum now runs over $z-1$
neighbors, excluding the site $i$. 

We note that this local action is identical as the action of 
an Anderson impurity model embedded in a sea of conduction electrons
described by a hybridization function $\Delta_{j} (\tau,\tau ')$.
We conclude that the objects $G_{j}^{(i)} (\omega_n )$ are 
related by a stochastic recursion relation, that involves 
solving Anderson impurity models with random on-site energies 
$\varepsilon_i$.

To make further progress, it is crucial to identify appropriate
 order parameters that can characterize different phases 
of the system and describe quantitatively
the approach to the transition.
In early work, it has already been stressed by Anderson \cite{anderson}
that a proper description of  disordered
systems should focus on distribution functions, 
and that {\em typical} rather than the average values 
should be associated with physical observables.
Our formalisms maps the original model onto an ensemble
of Anderson impurity models, and  its low energy behavior
is   naturally  described in terms of the distribution function
of the corresponding local density of states (DOS), defined as   
$\rho_j  =-Im G_{j} (0 )$  \cite{cavity}. From this distribution
we can extract    the typical DOS
$
\rho_{typ}=\exp\{ <\ln \rho >\}, 
$
which is a  natural order for the metal insulator transition.

On the metallic side of the transition,  the distribution
function of a second quantity, the local quasiparticle (QP)
weight, which is obtained from the Greens functions as
$q_j= {\partial \over {\partial  \omega} 
}Re[{G_j^{-1} - \Delta_j}]|_{\omega=0}$, 
is necessary to characterized  the low energy behavior near
the transition. Important information is obtained from
the typical value of   the  random variable $q_j$, 
 defined as
$
q_{typ}=\exp\{<\ln q_j >\},
$
which emerges as a natural order parameter from previous
studies of the Mott transition.

It is also useful to consider  the average
{\em quasiparticle} (QP)  density of states 
%
%\begin{equation}
$
\rho_{QP}= <\rho_j /q_j >.
$
This object  is very important for thermodynamics, since it is
directly related to quantities such as the specific heat
coefficient $\gamma =C/T$, or the local spin susceptibility $\chi_{loc}$. 

It is instructive to discuss the behavior of these order parameters in
the previously studied limiting cases.  In the limit of large lattice
coordination spatial fluctuations of the bath function $\Delta_i
(\omega_n )$ are unimportant, and there is no qualitative difference
between typical and average quantities.  In the Mott insulating phase
there is a gap in the density of states, while there is a finite
density of states on the metallic side of the transition.  As the MIT
is approached from the metallic side, $\rho_{typ}$ remains finite, but
$ q_{typ}$ is found \cite{dinfdis} to linearly go to zero.

Another well studied limit is that of noninteracting electrons on the
Bethe lattice, which is known \cite{anderson,abouchacra,efetov,mirlin}
to display an Anderson transition.  In the Anderson insulator phase
the local density of states has strong spatial fluctuations, few sites
with discrete bound states near the Fermi level have large density of
states while the density of states in most of the sites is zero.  The
average DOS is {\em finite} both in the insulating and in the metallic
phase, and is non critical at the transition.  Similarly, by
definition $ q_{typ}=1$ in this noninteracting limit, so it also
remains non critical.  On the other hand, the typical density of
states $\rho_{typ}$ is finite in the metal and zero in the Anderson
insulator.  This quantity is critical, and is found to 
vanishes exponentially \cite{bethe} with the distance to  the transition.

Equation \ref{eq3} is a system stochastic equations, i.e.  they depend
on the realization of the random variables describing the disorder.
To calculate the probability distributions of $\rho_j$ and $q_j$ we
use a simulation approach, where the probability distribution for the
stochastic quantity $G_{j}^{(i)} (\omega_n )$ is sampled from an
ensemble of N sites, as originally suggested by Abou-Chacra et al.
\cite{abouchacra}. To solve Anderson impurity models for given bath
functions $\Delta_{j} (\tau,\tau ')$ we use the slave boson (SB)
mean-field theory \cite{slaveboson} \cite{zimanyi}, which is known to
be qualitatively and even quantitatively correct at low temperature
and at low energies.

We now discuss our results for the nontrivial situation where both the
disorder and the interactions are present.  We consider a $z=3$ Bethe
lattice, in the limit of infinite on-site repulsion $U$ at $T=0$ and
fixed filling $n=0.3$, in the presence of a uniform distribution of random
site energies $\varepsilon_i$ of width $W$ (following the notation of
Ref. \cite{abouchacra}, $W$ is measured units of the hopping element
$t$). 
 We begin by concentrating on the evolution of the probability
distribution of the local quasiparticle weights $q_i$, as the disorder
is increased.  The sites with $q_i \ll 1$ represent \cite{msb,dinfdis}
disorder-induced local magnetic moments, and as such will dominate the
thermodynamic response (see the definition of $\rho_{QP}$).  For weak
disorder we expect relatively few local moments and the quasiparticle
weight distribution is peaked at a finite value.  As the disorder is
increased, the distribution of $q_j$-s broadens.  At a critical value
of the disorder $W_{nfl}$, a transition to a NFL metallic state takes
place.  To illustrate this behavior we display the integrated
distribution of the variable $q$, $n(q)$ for different values of
disorder in Fig. 1(a).  If $n(q)\sim q^\alpha$, as $q\rightarrow 0$,
and $\alpha \le 1$, then $P(q)\rightarrow +\infty$ in this
limit. Since the local Kondo temperatures $T_K^{(i)}\sim q_i$
\cite{dinfdis}, this behavior reflects a singular distribution of
Kondo temperatures.  As a result, we immediately obtain {\em non-Fermi
liquid} (NFL) behavior \cite{bf,dkk,miranda} with diverging $\gamma$ and
$\chi_{loc}$ 
\begin{figure}
\epsfxsize=3.2in \epsfbox{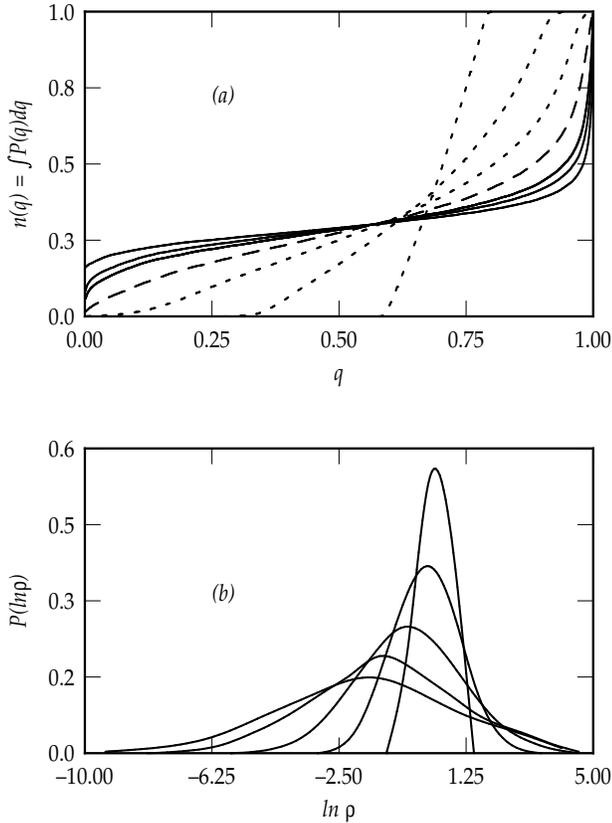}
\caption{Evolution of probability distributions for
{\em interacting} electrons as a function of disorder at $T=0$: (a)
integrated distribution for local quasiparticle weights (local Kondo
temperatures).  Results are presented for $W=1,3,5$ (dotted lines),
$W=7$ (dashed line), and $W=9,10,11$ (full lines).  The transition to
the NFL regime is signaled by the divergence of the {\em slope} of
$n(q)$ at $q=0$.  (b) The evolution of the local DOS distribution is
presented by plotting $P(\ln \rho )$ for $W=3,5,7,9,10$. We find that
the {\em maximum}, i. e. $<\ln\rho >$ shifts, as the transition is
approached. Note also the extremely large {\em width} of the
distribution, so that $\rho$ now spans many orders of magnitude. }
\end{figure}
at $T=0$.  As we can see, there is a well defined value
of disorder $W_{nfl}\sim 7$, beyond which the slope of $n(q)$ at
$q=0$ diverges, and we enter the NFL phase.  It is worth mentioning
that a similar transition to a NFL metal, well before the MIT, has
been found from the field-theoretical approaches in $2+\varepsilon$
dimensions \cite{fink,ckl,bk,ted}.  In the NFL phase the
thermodynamics is dominated by disorder-induced local moments.
The probability distribution of the second order parameter $\rho$,
$P(\ln \rho )$ , for different values of the disorder strength is
shown in Fig. 1(b).  Notice that not only the width, but also the
maximum of the distribution shifts with disorder, a behavior
reminiscent of an ordinary Anderson transition.  
The typical DOS is
strongly depressed at strong disorder.  This behavior is even more
clearly seen if we plot the DOS averages at the Fermi energy as a
function of disorder, as presented in Fig. 2(b).  The typical DOS
decreases in a clearly {\em linear fashion}, as the transition at
$W=W_c\approx 11$ is approached. 
\begin{figure}
\epsfxsize=3.2in \epsfbox{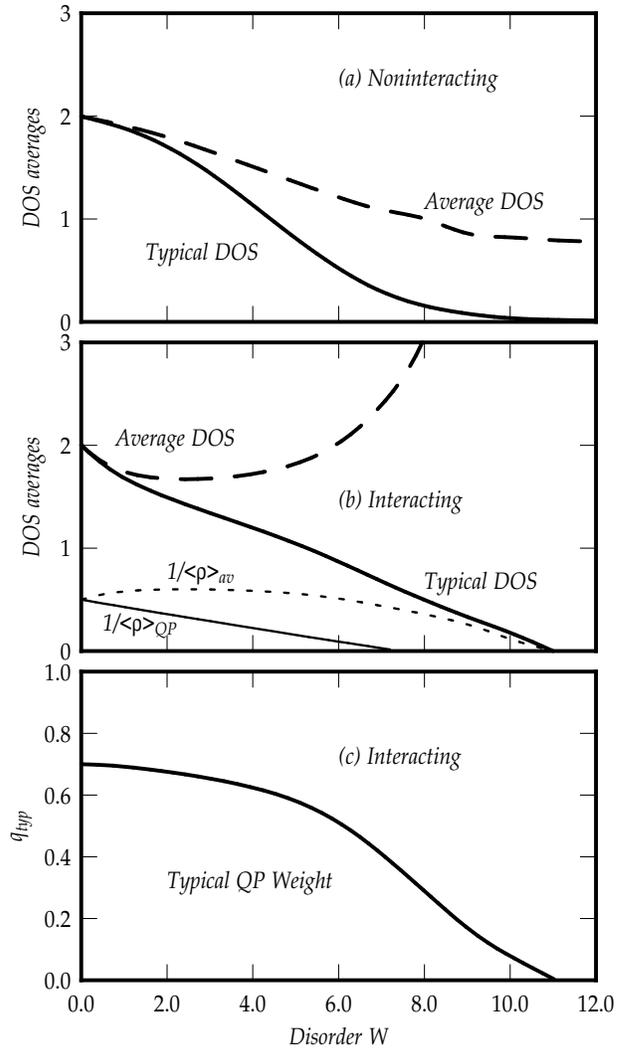}
\caption{Order parameters as functions of the disorder
strength $W$.  In the noninteracting limit (a), the typical DOS vanishes
exponentially with disorder, while the average DOS is non critical.
When interactions are present (b), the typical DOS decreases {\em
linearly} with disorder, while at the same time the average one {\em
diverges}.  The divergence is clearly seen by plotting $1/<\rho >_{ab}$
(dotted line), which vanishes {\em linearly} as the critical disorder is
approached.  Both quantities are found to be critical at $W=W_c\approx
11$.  Also shown is $1/<\rho >_{QP}$ (thin full line), which vanishes at
$W=W_{nfl}\approx 7$. Finally, we show in (c) the critical behavior of
the typical QP weight, which also vanishes linearly at $W=W_c$,
similarly as in a Mott transition} 
\end{figure}
This should be contrasted
\cite{transition} to the $U=0$ Anderson transition, where we find (see
Fig. 2(a)) the decrease to be exponential in agreement with analytical
results \cite{mirlin}.  We mention that at least in the noninteracting
limit \cite{mirlin}, the typical DOS behaves in a fashion which is
qualitatively identical to that of the diffusion coefficient.  Having
this in mind, one is tempted to interpret our results as indicating
linear behavior of the conductivity near the transition, as found
experimentally in many ``compensated'' systems.
Even more dramatic is the behavior of the average DOS 
which is non-critical both near a conventional $U=0$ Anderson
transition, and near a clean Mott transition.
This quantity is found to {\em diverge} at the same 
value of disorder where the typical DOS vanishes. 
The fact that we indeed have the divergence, is further confirmed 
by plotting $1/<\rho >_{av}$ as a function of disorder, as shown 
by a dotted line in Fig.2(b). This quantity vanishes linearly
at the same critical disorder $W=W_c\approx 11$.
In the same figure 
we exhibit the divergence of    the QP DOS,
at the transition to the NFL phase.
Finally, we consider the behavior of $q_{typ}$, which is also found to 
 vanish linearly at $W=W_c$, similarly as in the case of the Mott
transition, but in contrast to the noninteracting scenario. Physically, 
this indicates that a {\em finite fraction} of electrons turn into 
strictly localized magnetic moments at the metal-insulator transition. 

To summarize, in this paper we have presented a new self-consistent
theory of disordered interacting electrons that can describe both
the Anderson and the Mott route to localization. In this approach,
the typical local  DOS and the typical local
resonance width  play the  a role of  order 
parameters, but the entire {\em probability distributions} are  needed 
to fully characterize the behavior of the system. Our equations
take a form of stochastic recursion relations for these quantities
that involves solving an ensemble of Anderson impurity models. 
As a specific application of this approach, we have considered
a large $U$ 
limit of the Hubbard model at a fixed electron
density, and investigated effects
induced by gradually turning on the disorder. We find that 
the  correlations effects produce dramatic modifications 
of the conventional Anderson scenario. At intermediate disorder, 
there is a transition to a non-Fermi liquid phase, characterized by
singular thermodynamics, but conventional transport.
At larger disorder a metal-insulator transition takes place.
This  is a {\em new type} of transition, 
having some of the features of both the Anderson and the Mott scenario.
Remarkably, the main features our treatment, a non Fermi liquid phase
before the metal insulator transition and a linearly vanishing
conductivity   are found in
compensated doped semiconductors.

Our framework suggest several research directions. One would like to
relate response functions that determine the transport coefficients to
the local order parameters, as was done in the non interacting case by
Efetov and Viehweger \cite{mirlin}.  Our calculations  should be extended to
to the vicinity of half filling where correlations effects should be
even more pronounced. This study could cast some light on the
different types of metal insulator transitions that occur in
compensated and uncompensated doped semiconductors. 

One of us (VD) acknowledges useful discussions with Sasha Finkelshtein, 
Lev Gorkov, E. Miranda, J. R. Schrieffer, and G. Thomas. 
VD was supported by the
National High Magnetic Field Laboratory at Florida State University.
GK was supported by NSF DMR 95-29138.


\begin{references}
\bibitem{bethe} This unusual exponential critical behavior is specific
to the Bethe lattice, in contrast to the usual power law dependence
expected to hold in finite dimensions.  However, detailed studies
\cite{mirlin} have shown that all the other qualitative features of an
Anderson transition are present even for the Bethe lattice, so we
believe that this model does contain all the crucial ingredients
required to investigate the interplay of localization and correlation
effects.  Interestingly, we find that interactions eliminate the
exponential behavior, so we expect our conclusions to be valid 
for general lattices.
\end{references}
\end{document}